\documentclass[11pt,twoside]{article}  
\usepackage{asp2006}
\usepackage{adassconf}

\begin{document}   

%
%

\paperID{O08.3}

%

\title{The Herschel Data Processing System --- HIPE and pipelines --- up and running since the start of the mission}

%
%
%
%
%

\markboth{Ott}{The Herschel Data Processing System}

%
%
%
%

\author{S.\ Ott\altaffilmark{1}}
\affil{Herschel Science Centre, European Space Agency}
\altaffiltext{1}{on behalf of
the Herschel Science Ground Segment Consortium\\
http://herschel.esac.esa.int/DpHipeContributors.shtml}

%

\contact{Stephan Ott}
\email{sott@rssd.esa.int}

%
%
%

\paindex{Ott, S.}

%



\begin{abstract}          
The Herschel Space Observatory is the fourth cornerstone mission in
the ESA science programme and performs photometry and spectroscopy
in the 55 - 672 micron range. The development of the Herschel Data
Processing System started in 2002 to support the data analysis for
Instrument Level Tests. The Herschel Data Processing System was used
for the pre-flight characterisation of the instruments, and during
various ground segment test campaigns. Following the successful
launch of Herschel 14th of May 2009 the Herschel Data Processing
System demonstrated its maturity when the first PACS preview
observation of M51 was processed within 30 minutes of reception of
the first science data after launch. Also the first HIFI
observations on DR21 were successfully reduced to high quality
spectra, followed by SPIRE observations on M66 and M74. A fast
turn-around cycle between data retrieval and the production of
science-ready products was demonstrated during the Herschel Science
Demonstration Phase Initial Results Workshop held 7~months after
launch, which is a clear proof that the system has reached a good
level of maturity.

We will summarise the scope, the management and development
methodology of the Herschel Data Processing system, present some key
software elements and give an overview about the current status and
future development milestones.
\end{abstract}

%
%

\section{Introduction}

The Herschel Space Observatory, the fourth cornerstone mission in
the ESA science programme, was successfully launched 14th of May
2009. With a 3.5~m Cassegrain telescope it is the largest space
telescope ever launched (Pilbratt et al., 2008). Herschel's three
instruments (HIFI, PACS and SPIRE) perform photometry and
spectroscopy in the 55 - 672 micron range and will deliver exciting
science for the astronomical community during at least three years
of routine observations (de Graauw et al., 2008, Poglitsch et al.,
2008, Griffin et al., 2008). One month after launch, on its way to
its operational orbit around L2, the Lagrange point located 1.5
million kilometres away from the Earth, the cryostat lid was opened,
and the first observational tests were conducted. Most of Herschel's
performance verification and science demonstration phase activities
have been completed for SPIRE and PACS, and both instruments will
commence routine operations before 2009 ends.

\section{Scope, Management and Development Methodology of the Herschel
Data Processing System}

The development of the Herschel Data Processing System started in
2002 to support the data analysis for Instrument Level Tests with
the final goal to provide an integrated, easy to use, well tested
and documented data processing system to the astronomical community
free of charge. Subsequently the Herschel Data Processing System was
used for the pre-flight characterisation of the instruments, and
exercised during various ground segment test campaigns. The system
combines for the first time data retrieval, pipeline execution and
scientific analysis in one single environment. All tools for data
reduction and analysis, e.g. also the expert applications for
'Instrument Calibration', 'Trend Analysis' and 'Quality Control'
systems are part of the Herschel Data Processing System. Therefore
the community has access to the same system as the instrument
experts. Also the Standard Product Generation (SPG) software which
automatically generate data products are a subset of the Herschel
Data Processing System.

The Herschel Data Processing system is coded in Java/Jython to be
license free and portable for different operating systems. Its
source code is freely available under the GNU lesser general public
license and is already accessible to Herschel Key Programme users.

Developing the Herschel Data Processing system is a major project,
with over 200 contributors and currently 60 full-time equivalents
working on calibration, coding, documentation, quality control,
testing and tutoring. The Herschel Science Centre (ESA), the
Instrument Control Centres (HIFI, PACS and SPIRE) and the NASA
Herschel Science Center (NHSC) jointly manage and contribute to the
Herschel Data Processing System.
\vspace{-1ex}
\section{Key Software Elements of the Herschel Data Processing System}
Installers are available for a variety of operating systems and
formal support is provided for Windows XP, Vista, Linux, Mac~OS~X
10.5 ("Leopard") for both user and developer versions of the
Herschel Data Processing System. Formal support of Windows~7 and
Mac~OS~X 10.6 ("Snow Leopard") will be provided in the future.

The Herschel Interactive Processing Environment (HIPE) was
introduced as the user friendly face of the Herschel Data Processing
System. It provides both a script driven, command line based
environment suited to developers' and experts' needs and a GUI based
end user-oriented environment. This interface is more data-centric
than language-centric, providing astronomers who are not experienced
in Java a state of the art interface to process Herschel data. It
also has the great advantage that the same framework can be used to
download, reprocess, analyse and compare data from all three
instruments in Herschel simultaneously.

HIPE is rich in features: Drag 'n Drop, command-line echoing, access
to functionality by menu, toolbar, pop-up menus, and keyboard
short-cuts. It allows to analyse the same data in different ways and
to compare similar types of data easily. It features a user
controllable layout, permitting to organise the views the way the
user wants it, add/remove them, and to dock and undock views. HIPE
allows the user to execute the official pipeline and to adapt it to
their own scientific needs within the same environment.

Another advanced feature of the Herschel Data Processing system is
the Product Access Layer, an interface to store, query and load
Herschel data in/from different types of specialised Products pools,
the documentation generation system, and the direct access to the
Herschel Science Archive, which allows users to retrieve the raw
data from the HSA directly in batch mode. The Product Access Layer
also permits to share data between collaborators.

A state of the art documentation system for the framework and all
instruments which includes context-sensitive search functionality is
provided as part of the Herschel Data Processing System.
\vspace{-1.0ex}
\section{Operational Use of the Herschel Data Processing System}
The data processing pipelines which automatically generate data
products are executed on the European Space Astronomy Centre (ESAC)
computing grid to produce Herschel Products to different reduction
levels: \vspace{-0.6ex}
\begin{description}
\item[Level 0 products] Raw telemetry data as measured by the
instrument. They might be minimally formatted\\[-3.6ex]
\item[Level 1 products] Detector readouts calibrated and converted
to physical units, in principle instrument and observatory
independent\\[-3.6ex]
\item[Level 2 products] Level-1 data further processed to absolutely
calibrated images, spectral cubes and spectra so that scientific
analysis can be performed
\end{description}
\vspace{-0.6ex} The data products are ingested into the Herschel
Science Archive (HSA) and made available to the data owners through
a user interface similar to those of existing ISO and XMM archives,
usually on the same day of reception of the data from the satellite.

Data quality control is performed for all Herschel scientific
observations by the observatory's Technical Assistants and
Instrument Calibration Scientists. This data quality control is a
combination of automatic screening and manual inspection. Quality
control reports are electronically distributed to experts and
usually takes a few days. A summary of these findings is contained
in the quality control summary that is made available as part of the
Herschel data products. \vspace{-0.8ex}
\section{Handling of Herschel data during its early operational days}
Herschel opened its eyes on 14 June 2009, precisely one month after
the launch from Kourou, as the cryocover, the cryostat lid, was
commanded to open. During the remainder of the operational day
Herschel carried out test observations labelled a 'sneak preview'.
Within 30 minutes following the reception of these data from the
first PACS observation of the whirlpool galaxy M51 a fantastic image
was generated within HIPE, followed by the operational pre-launch
pipeline. Also spectra showing CO and H$_2$O lines of HIFI's first
light observation of the star forming region DR21 were generated
both in the interactive environment and in the standard pipeline.
Finally SPIRE's first light observations of M66 and M74 were
successfully handled by both HIPE and pipeline. Other early results
were the handling of PACS imaging spectroscopy on the Cat's Eye'
nebula NGC6543 and of the PACS/SPIRE parallel mode that revealed new
details in the Milky Way.

\begin{figure}[t]
\plottwo{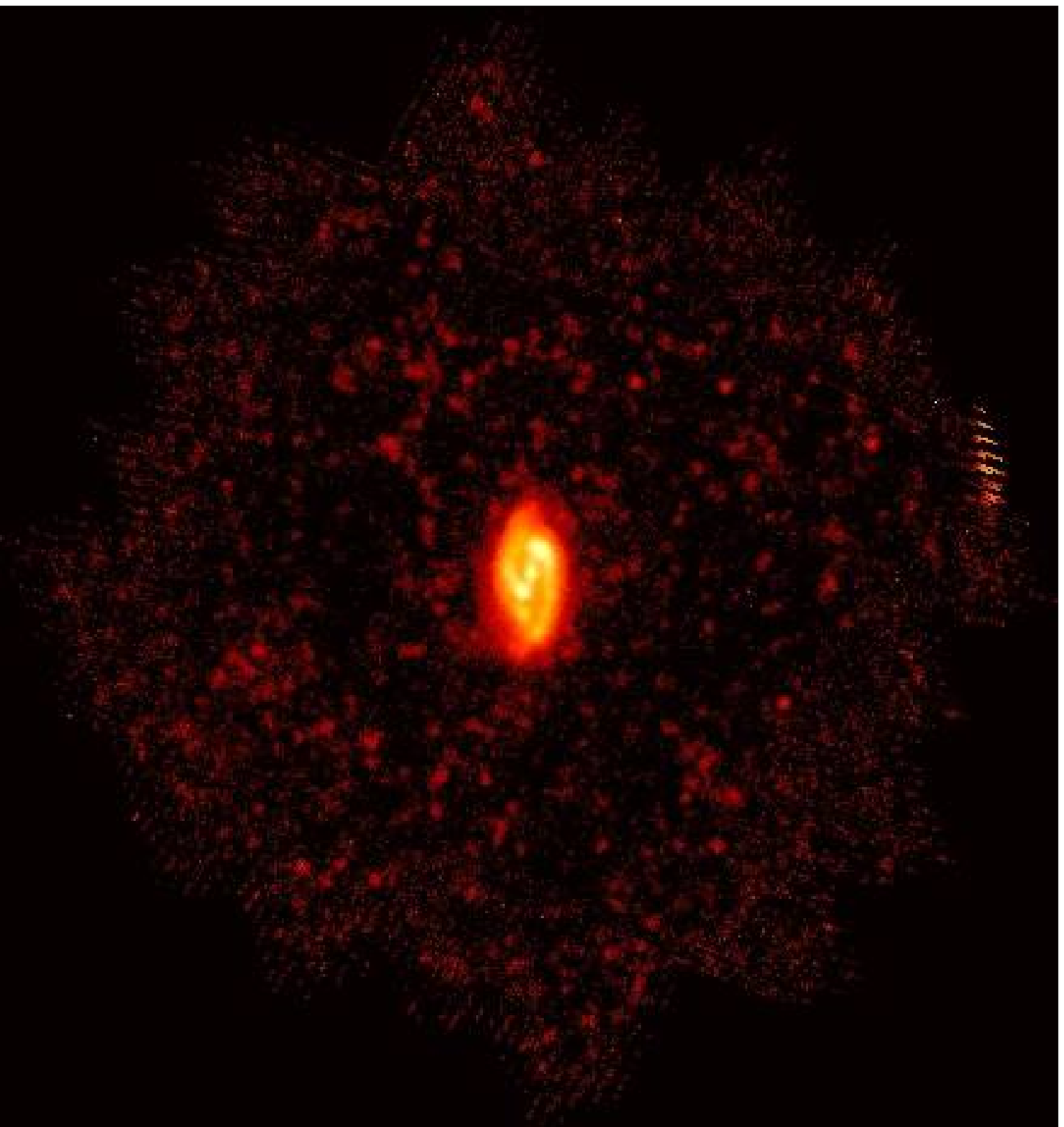}{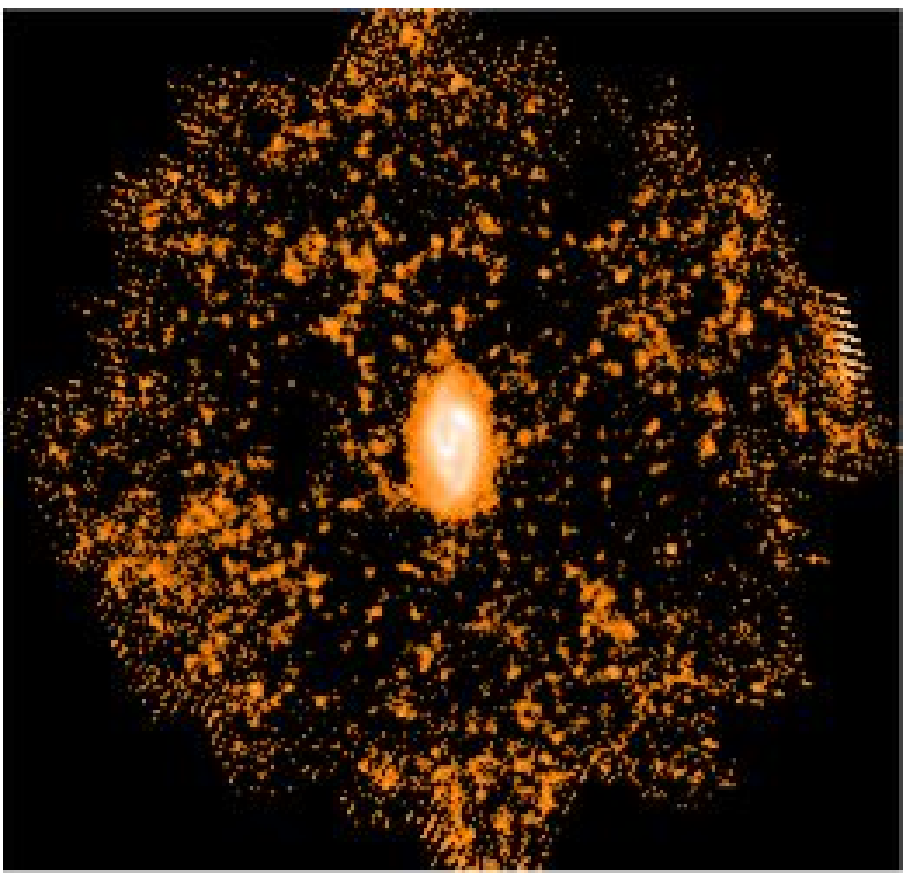}\caption{SPIRE
first light observation on M66 reduced by HIPE (left) and the
current operational pipeline (right).} \label{O08.3-fig-1}
\end{figure}

\vspace{-1.0ex}
\section{Current Status and future Development Milestones}
HIPE 1.1 (Performance Verification Phase version) was made available
to the Herschel Key Program teams. This version was also used to
perform the first bulk reprocessing exercise of Herschel data. The
first Science Demonstration Phase observations were conducted in
September 2009 and data products were provided to the observers
using an early version of HIPE 1.2. As of end 2009 HIPE 1.2 (Science
Demonstration Phase version) is operational. HIPE 2.0 (Science
Routine Phase version) is currently under testing. An early version
of HIPE 2.0 was made available to the participants of the Herschel
Science Demonstration Phase Data Processing Workshop. HIPE 2.0 will
be made available to the Herschel community early next year. It is
foreseen that future HIPE versions will be released regularly;
during the following year around each three months.


\vspace{-0.7ex}
\begin{references}
\reference Pilbratt, G.~L.\ et al.\ 2008, Proc.\ SPIE, 7010 02

\reference de Graauw, M.~W.~M.\ et al.\ 2008, Proc.\ SPIE, 7010 04

\reference Poglitsch, A.\ et al.\ 2008, Proc.\ SPIE, 7010 05

\reference Griffin, M.~J.\ et al.\ 2008, Proc.\ SPIE, 7010 06

\reference Herschel's 'sneak preview', 'first-light' and PACS/SPIRE
parallel mode results,\\
http://herschel.esac.esa.int/SneakPreview.shtml\\
http://herschel.esac.esa.int/FirstLight.shtml\\
http://herschel.esac.esa.int/FirstParallelModeImages.shtml

\reference Herschel Science Demonstration Phase Initial Results
Workshop, \\http://herschel.esac.esa.int/SDP\_IR\_wkshop.shtml
\end{references}
\end{document}